\documentclass[preprint,prc,aps,superscriptaddress,nofootinbib,showpacs]{revtex4}
\usepackage[dvips]{graphicx}
\usepackage{amsfonts, color, float, bbm, amsmath}
\usepackage{amssymb}

\def\lp {\left( }
\def\rp {\right) }
\def\lb {\left[ }
\def\rb {\right] }
\def\lc {\left\{ }
\def\rc {\right\} }
\def\ra {\rangle }
\def\la {\langle }
\def\ni {\noindent}
\def\nn {\nonumber}
\def\vs {\vspace}
\def\rar {\rightarrow}

\def\beq{\begin{equation}}
\def\eeq{\end{equation}}
\def\bea{\begin{eqnarray}}
\def\eea{\end{eqnarray}}

\def\dr {\partial }

\def\lga {\simeq^{\!\!\!\!\!\!\!^{lga}} }

\def\cE {{\cal{E}}}
\def\cI {{\cal{I}}}

\def\cN {{\cal{N}}}
\def\cO{{\cal{O}}}
\def\cT {{\cal{T}}}

\def\d{\delta}

\def\e{\epsilon}
\def\g{\gamma}

\def\l {\lambda}
\def\m{\mu}
\def\n{\nu}

\def\O {\Omega}
\def\p {\pi}
\def\r{\rho}
\def\s{\sigma}

\def\ub {\bar u}

\def\sp {\!+\!}
\def\sm {\!-\!}
\def\st {\!\times \!}
\def\cd {\!\cdot\!}

\def\br {\mbox{\boldmath $r$}}
\def\bR {\mbox{\boldmath $R$}}
\def\bro {\mbox{\boldmath $\rho$}}

\def\bsig {\mbox{\boldmath $\sigma$}}
\def\btau {\mbox{\boldmath $\tau$}}
\def\bL {\mbox{\boldmath $L$}}
\def\bnb {\mbox{\boldmath $\nabla$}}
\def\bO {\mbox{\boldmath $\Omega$}}
\def\bp {\mbox{\boldmath $p$}}
\def\bP {\mbox{\boldmath $P$}}

\def\bq {\mbox{\boldmath $q$}}
\def\bQ {\mbox{\boldmath $Q$}}

\begin{document}

\title{NUCLEON-NUCLEON POTENTIAL:\\
DRIFT EFFECTS}

\author{M. R. Robilotta}
\email[]{robilotta@if.usp.br}
\affiliation
{Instituto de F\'{\i}sica, Universidade de S\~{a}o Paulo,\\
C.P. 66318, 05315-970, S\~{a}o Paulo, SP, Brazil.}

\date{\today}

\begin{abstract}
In the rest frame of a many-body system, used in the calculation of its static and 
scattering properties, the center of mass of a two-body subsystem is allowed to drift.
We show, in a model independent way, that drift corrections to the nucleon-nucleon  
potential are relatively large and arise from both one- and two-pion exchange processes.
As far as chiral symmetry is concerned, corrections to these processes begin 
respectively at $\cO(q^2)$ and $\cO(q^4)$.
The two-pion exchange interaction also yields a new spin structure, that promotes the presence 
of $P$ waves in trinuclei and is associated with profile functions which do not coincide 
with neither central nor spin-orbit ones.
In principle, the new spin terms should be smaller than the $\cO(q^3)$ spin-orbit components.
However, in the isospin even channel, a large contribution reverts this expectation 
and gives rise to the prediction of important drift effects.

\end{abstract}

\maketitle


\section{introduction}

This work is motivated by a private question posed by Alejandro Kievsky some years ago, 
concerning the possibility of novel forms of spin dependence in the nucleon-nucleon interaction,
when one is not in the center of mass frame of the two-body system.
In the study of static and scattering properties of many-body nuclei, calculations are 
performed in the rest frame of the larger system and a two-body subsystem is allowed to drift.
This picture led him to introduce a phenomenological three body force\cite{K}, 
which improved the description  of the $N \sm d$ vector analyzing power $A_y$.

Nowadays, the outer layers of the $NN$ interaction, represented by one-pion and two-pion exchange 
potentials ($ OPEP$ and $TPEP$), are set in solid foundations due to the use of chiral symmetry.
Nuclear processes are dominated by the light quarks $u$ and $d$, and one is not far from the 
massless limit, in which QCD becomes invariant under both isospin and chiral 
$SU(2)\times SU(2)$ transformations.
Chiral symmetry is realized in the Nambu-Goldstone mode and the QCD vacuum can bear collective 
excitations, identified as pions.
A suitably formulated chiral perturbation theory (ChPT) allows deviations from the massless 
limit to be treated systematically\cite{ChPT}.
As low-energy QCD calculations are prevented by its non-Abelian character,
in practice one works with chiral effective theories, in which elementary nucleons interact 
by exchanging pions.

In chiral perturbation one uses a typical scale $q<<1$ GeV, set by either pion four-momenta 
or nucleon three-momenta.
The leading term\cite{Bira} in the $NN$ interaction is the $OPEP$, at $\cO(q^0)$.
The $TPEP$ begins at $\cO(q^2)$ and two independent expansions up to $\cO(q^4)$ are 
presently available.
One of them, based on heavy baryon ChPT\cite{HB}, uses non-relativistic lagrangians from the 
very beginning and the inverse of the nucleon mass as an expansion parameter.
The other one, produced recently by our group\cite{HR,HRR}, is based on relativistic expressions, 
written in terms of observable coefficients and covariant loop integrals.
The use of a relativistic language frees one from particular reference frames and allows 
a straightforward treatment of two-body interactions in which the center of mass  
is able to move.
Here, we rely on our previous work in order to derive the drift contributions to the 
$NN$ potential.
For the sake of definiteness, we stay in the realm of  three-body nuclei, but results 
can be easily generalized to larger systems.

Our presentation is organized as follows.
In section II, we review the dynamical role of two-body interactions in trinuclei.
This sets the stage for the derivation of drift interactions, which is performed in section III.
Results are summarized in section IV, whereas technical issues, concerning kinematics and 
spin operators, are left to appendices.

\section{dynamics}

The interactions of a three-nucleon system in momentum space are represented by the operator $W$, 
defined by\cite{Y,CDR}  

\beq
\la \bp'_1, \bp'_2, \bp'_3 \,| \hat{W} |\, \bp_1, \bp_2, \bp_3 \ra = - (2\p)^3 \, 
\d^3(\bp'_1 \sp \bp'_2 \sp \bp'_3 \sm \bp_1 \sm \bp_2 \sm \bp_3) \; 
\bar{t}_3 (\bp'_1, \bp'_2, \bp'_3, \bp_1, \bp_2, \bp_3)  \;,
\label{2.1}
\eeq

\ni 
where $\bar{t}_3$ is the proper part of the non-relativistic three-body transition matrix.
In configuration space, the position of nucleon $i$ is described by $\br_i$ 
and one uses the Jacobi variables

\beq
\bR = (\br_1 \sp \br_2 \sp \br_3) /3\,, \;\;\;\;\;\;\;\;
\br = \br_2 \sm \br_1 \,, \;\;\;\;\;\;\;\;
\bro = (2\, \br_3 \sm \br_1 \sm \br_2)/\sqrt{3}\,,
\label{2.2}
\eeq

\ni
which correspond to 

\beq
\bp_1 =  \frac{\bP}{3} - \bp_r  -  \frac{\bp_\r}{\sqrt{3}} \,, \;\;\;\;\;\;\;\;
\bp_2 =  \frac{\bP}{3} + \bp_r - \frac{\bp_\r}{\sqrt{3}}  \,, \;\;\;\;\;\;\;\;
\bp_3 =  \frac{\bP}{3} + \frac{2 \,\bp_\r}{\sqrt{3}} \,.
\label{2.3}
\eeq

The Schr\"odinger equation for the internal degrees of freedom is obtained by using 
$\bP = \bP' = 0$ and given by

\beq
 \lb - \frac{1}{m}\,\bnb_{r'}^2 - \frac{1}{m}\,\bnb_{\r'}^2 - \e \, \rb \psi(\br', \bro')
= - \lb \frac{\sqrt{3}}{2}\rb^3 \int d \br \, d\bro \; W(\br', \bro'; \br, \bro) 
\; \psi (\br, \bro) \;,
\label{2.4}
\eeq

\ni
with 

\bea
&& W(\br', \bro'; \br, \bro)  = -\, \frac{1}{(2\p)^{12}}\; \lb \frac{2}{\sqrt{3}} \rb^6
\int d\bQ_r \, d\bQ_\r \, d\bq_r \, d\bq_\r \;
\nn\\[2mm]
&& \;\;\;\;\; \;\;\;\;\; \times
e^{i \lb \bQ_r \cdot (\br' \sm \br) + \; \bQ_\r \cdot (\bro' \sm \bro)  
+ \bq_r \cdot (\br' \sp \br) /2 + \bq_\r \cdot (\bro' \sp \bro) /2   \rb}
\;\; \bar{t}_3 (\bQ_r, \bQ_\r, \bq_r, \bq_\r)  \;,
\label{2.5}
\eea

\ni
and $\bQ_i = (\bp_i' \sp \bp_i)/2$ and  $\bq_i = (\bp_i' \sm \bp_i)$, for $i=(r, \rho)$.

In this work we are interested in describing two-body interactions between nucleons 
$1$ and $2$ and note that the conservation of $\bp_3$ implies $\bq_\r = 0 $.
We write  

\beq
\bar{t}_3 (\bp'_1, \bp'_2, \bp'_3; \bp_1, \bp_2, \bp_3) 
= (2\p)^3 \, \lb \frac{\sqrt{3}}{2} \rb^3 \, \d^3(\bq_\r) \; 
\bar{t}_2(\bQ_\r, \bQ_r, \bq_r) \;,
\label{2.6}
\eeq 

\ni
where $\bar{t}_2$ is the two-body $t$-matrix, and the corresponding potential becomes

\bea
&& W_2(\br', \bro'; \br, \bro) = -\,\frac{1}{(2\p)^9}\, \lb \frac{2}{\sqrt{3}} \rb^3  \,
\int d\bQ_r \, d\bQ_\r \, d\bq_r  \;
\nn\\[2mm]
&& \;\;\;\;\; \;\;\;\;\; \times
e^{-i \lb \bQ_r \cdot (\br' \sm \br) + \; \bQ_\r \cdot (\bro' \sm \bro)  
+ \bq_r \cdot (\br' \sp \br) /2 \rb}
\;\; \bar{t}_2(\bQ_r, \bQ_\r, \bq_r )  \;.
\label{2.7}
\eea

In isospin space, the amplitude $\bar{t}_2$ reads  

\beq
\bar{t}_2 =  t^+ + \btau^{(1)} \!\cdot \btau^{(2)} \, t^- \;.
\label{2.8}
\eeq

The usual spin decomposition is obtained by going to the center of mass frame 
of the two-body system, where one finds

\beq
\left. t_2^\pm \rb _{cm}  
=  t_C^\pm  + \frac{\bO_{LS}}{m^2} \, t_{LS}^\pm 
+ \frac{\bO_{SS}}{m^2}\, t_{SS}^\pm + \frac{\bO_{T}}{m^2}\, t_{T}^\pm 
+ \frac{\bO_{Q}}{m^4}\, t_{Q}^\pm \;,
\label{2.9}
\eeq

\ni
with two-component operators defined by

\bea
&& \bO_{LS} = i \, (\bsig^{(1)}\sp \bsig^{(2)}) \cd \bq_r \! \times \! \bQ_r /2 \;,
\label{2.10}\\[2mm]
&& \bO_{SS} = \bq_r^2\, \bsig^{(1)} \! \cd \! \bsig^{(2)} \;,
\label{2.11}\\[2mm]
&& \bO_{T} = - \bq_r^2\,(3\; \bsig^{(1)} \! \cd \! \hat{\bq_r}\; \bsig^{(2)} \!\cd \! \hat{\bq_r} 
- \bsig^{(1)} \cd \bsig^{(2)}) \;,
\label{2.12}\\[2mm]
&& \bO_{Q} = 4\, \bsig^{(1)}\cd \bq_r \! \times \! \bQ_r  \;  
\bsig^{(2)}\cd \bq_r \! \times \! \bQ_r  \;.
\label{2.13}
\eea

In this formulation, the two-body interaction does not depend on $\bQ_\r$ and is 
completely decoupled from the larger system it is immersed in.
The Fourier transform of this result produces the configuration space potential, given by 

\bea
&& W_2(\br', \bro'; \br, \bro)= \d^3(\br' \sm \br)\, \d^3(\bro' \sm \bro)
\lb \frac{2}{\sqrt{3}}\rb^3\, \left. V(\br)^\pm \rb_{cm}  \;, 
\nn\\[2mm]
&& \left. V(r)^\pm \rb_{cm}  = V_C^\pm + V_{LS}^\pm \, \O_{LS} + V_{SS}^\pm \, \O_{SS} 
+ V_T^\pm \, \O_T\;, 
\label{2.14}
\eea

\ni
where we have kept only local and spin-orbit contributions and the spin operators read

\bea
&& \O_{LS} = \bL \cd (\bsig^{(1)}\sp \bsig^{(2)})/2 \;,
\label{2.15}\\[2mm]
&& \O_{SS} = \bsig^{(1)}\cd \bsig^{(2)}\;,
\label{2.16}\\[2mm]
&& \O_T = 3\, \bsig^{(1)}\cd \hat{\br} \; \bsig^{(2)}\cd \hat{\br} - \bsig^{(1)}\cd \bsig^{(2)} \;.
\label{2.17}
\eea

The radial functions are given by 

\bea
V_{C}^{\pm}(r) &=&  U_{C}^{\pm}(x)\,,
\label{2.18}\\[2mm]
V_{LS}^{\pm}(r) &=&  \frac{\mu^2}{m^2}\,\frac{1}{x}\, \frac{d}{d x}\,U_{LS}^{\pm}(x)\,,
\label{2.19}\\[2mm]
V_{SS}^{\pm}(r) &=& - \frac{\mu^2}{m^2}\left[\frac{d^2}{d x^2}
+\frac{2}{x}\,\frac{d}{d x}\right]\,U_{SS}^{\pm}(x)\,,
\label{2.20}\\[2mm]
V_{T}^{\pm}(r) &=&  \frac{\mu^2}{m^2}\left[\frac{d^2}{d x^2}
-\frac{1}{x}\,\frac{d}{d  x}\right]\,U_{T}^{\pm}(x)\,,
\label{2.21}
\eea

\ni
with $x=\m r$ and

\beq
U_I^\pm(x) =  - \int \frac{d^3q}{(2\pi)^3}\, e^{i\,\bq \cdot \br }\; t_I^\pm(q) \,,
\qquad\qquad I=\{C,LS,SS,T\}\,.
\label{2.22}
\eeq 

As we discuss in the sequence, the fact that the two-body CM is allowed to drift 
gives rise to extra interaction operators in the potential.

\section{drift terms}

Corrections to the $NN$ potential due to the motion of the CM are derived by evaluating $\cT$,
the covariant $t$-matrix  for the on-shell process $N(p_1)\,N(p_2) \rar N(p'_1)\,N(p'_2)$, 
and writing the result in terms of two-component spinors, using the expressions of appendix B. 
This gives rise to an amplitude expanded in terms of Pauli spin operators.
Dividing it by the factor $4mE$ present in the relativistic normalization, 
one obtains the amplitude $\bar{t}_2$, which is to be fed into eq.(\ref{2.7}).   
In this work we concentrate on contributions from processes due to the exchanges 
of one and two pions.

The transformation of a $t$-matrix into a potential to be used in a dynamical equation 
is not trivial and depends on a number of important conventions.
These range from the very nature of of the equation adopted to tacit assumptions 
concerning the off-shell behavior of the potential.
The latter class of effects appears as corrections to leading order effects 
and was discussed in a comprehensive paper by Friar\cite{Fr}.
We here stick to the conventions used long ago by Partovi and Lomon\cite{PL,RR94}.

\vs{3mm}
\ni
{\bf $\bullet$ OPEP}

The covariant amplitude for on-shell nucleons reads

\beq
\cT = \btau^{(1)} \!\cdot \btau^{(2)} \; \frac{g_A^2 m^2}{f_\p^2}\;
\frac{1}{q^2 -\m^2}\; [\ub\, \g_5\, u]^{(1)} \, [\ub\, \g_5\, u]^{(2)} \;,
\label{3.1}
\eeq

\ni
where $g_A$, $f_\p$, $\m$, $m$ are respectively the axial and pion decay constants, 
the pion and nucleon masses.
Using eq.(\ref{a.8}) for the momentum $q$ and eq.(\ref{b.4}) for the spinor matrix element, 
one finds the two-component amplitude

\bea
\cT &\!=\!& \btau^{(1)} \!\cdot \btau^{(2)} \; \frac{g_A^2 m^2}{f_\p^2}\;
\frac{1}{(\bq_r^2 \sp \m^2) \sm 4\, (\bq_r \cd \bQ_\r)^2 /3 \cE^2}\;
\nn\\[2mm]
&& \cN^2 \; \lc [m \sp \cE/2 + 2\, \bQ_r \cd \bQ_\r / \sqrt{3}\, \cE ] \, \bsig^{(1)}\cd \bq_r
- 2\, \bq_r \cd \bQ_\r \, \bsig^{(1)}\cd (\bQ_r \sp \bQ_\r / \sqrt{3}) / \sqrt{3}\, \cE \rc
\nn\\[2mm]
&& \lc [m \sp \cE/2 - 2\, \bQ_r \cd \bQ_\r / \sqrt{3}\, \cE ] \, \bsig^{(2)}\cd \bq_r
+ 2\, \bq_r \cd \bQ_\r \, \bsig^{(2)}\cd (\bQ_r \sm \bQ_\r / \sqrt{3}) / \sqrt{3}\, \cE \rc \;,
\label{3.2}
\eea

\ni
where $\cN^2$ is a normalization factor, 

\beq
\cN^2 = \lc \lb (m\sp \cE/2)^2 \sp [4(\bQ_r \cd \bQ_\r)^2 \sm (\bq_r \cd \bQ_\r)^2]/3 \cE^2 \rb^2
- 16 (m \sp \cE/2)^2 (\bQ_r \cd \bQ_\r)^2 / 3 \cE^2 \rc^{-1/2}
\label{3.3}
\eeq

\ni
and $\cE$ is the total energy 
of the two-nucleon system, determined by the condition

\beq
\cE^4 - 4 (m^2 \sp \bq_r^2 \sp 4 \bQ_r^2 \sp \bQ_\r^2/3) \, \cE^2
+ (4/3) \,  \lb (\bq_r \cd \bQ_\r)^2 + 4\, (\bQ_r \cd \bQ_\r)^2 \rb = 0 \;.  
\label{3.4}
\eeq

This $t$-matrix is fully relativistic and contains no approximations.
All its terms involving the variable $\bQ_\r$ vanish in the rest frame of the two-body system
and therefore can be interpreted as drift effects.
With the provisos discussed in ref.\cite{Fr}, it could already be used as input 
into a momentum space dynamical equation.
Alternatively, in the framework of chiral perturbation theory, one might wish to
rewrite it as a power series, truncated at a given order. 

In configuration space, the variables $\bQ_r$ and $\bQ_\r$ correspond to non-local operators
and are usually associated with gradients acting on the wave function.
In order to restrict the corresponding complications to a minimum, 
we remain in the limited scope of eq.(\ref{2.4}) and keep only terms linear in these momenta.
This procedure is referred to as the {\em linear gradient approximation}.

Within this approximation, the $OPEP$ retains its usual local form, given by 

\beq
t_{SS}^- = - t_T^- \lga \frac{g_A^2 m^2}{12\,f_\p^2}\; \frac{1}{\bq_r^2 + \m^2} 
+ \mathrm{local \; corrections}\;.
\label{3.5}
\eeq

\vs{3mm}
\ni
{\bf $\bullet$ TPEP}

Quite generally, for each isospin channel, the spin content of the $TPEP$ is given by\cite{HR}

\bea
\cT^\pm  \!\!\!\!\! && =  [\ub\, u]^{(1)} \,  [\ub\,u]^{(2)} \, (\cI_{DD}^\pm)
-\frac{i}{2m}\,  [\ub\, u]^{(1)}\, [\ub \, \s_{\m\l}(p' \sm p)^\m \, u]^{(2)}\, (\cI_{DB}^\pm)^\l 
\nn\\[2mm]
&& -\, \frac{i}{2m}\,  [\ub \, \s_{\m\l}(p' \sm p)^\m \, u]^{(1)}\,  
[\ub\, u]^{(2)}\, (\cI_{BD}^\pm)^\l  
\nn\\[2mm]
&& -\, \frac{1}{4m^2}\,  [\ub \, \s_{\m\l}(p' \sm p)^\m \, u]^{(1)}\,  
[\ub \, \s_{\n\r}(p' \sm p)^\n \, u]^{(2)}\, 
(\cI_{BB}^\pm)^{\l\r} \;,
\label{3.6}
\eea

\ni
where the functions $\cI$ involve loop integrals and have a Lorentz structure realized 
in terms of the kinematical variables $W$, $z$ and $q$, defined in appendix A.
Terms proportional to $q$ do not contribute for on-shell nucleons and we have

\bea
&& (\cI_{DB}^\pm)^\l = \frac{W^\l}{2m} \, \cI_{DB}^{(w)\pm} 
+ \frac{z^\l}{2m}\, \cI_{DB}^{(z)\pm}\;,
\label{3.7}\\[2mm]
&& (\cI_{BD}^\pm)^\l = \frac{W^\l}{2m} \, \cI_{DB}^{(w)\pm} 
- \frac{z^\l}{2m}\, \cI_{DB}^{(z)\pm}\;,
\label{3.8}\\[2mm]
&& (\cI_{BB}^\pm)^{\l\r} = g^{\l\r} \, \cI_{BB}^{(g)\pm} 
+ \frac{W^\l W^\r}{4m^2} \,\cI_{BB}^{(w)\pm}
+ \frac{z^\l z^\r}{4m^2} \,\cI_{BB}^{(z)\pm}\;.
\label{3.9}
\eea

The amplitudes $\cI$ were explicitly calculated in ref.\cite{HR},
as functions of the invariants $W^2$, $z^2$ and $q^2$,
and the two-pion exchange interaction is described by 

\bea
&& \cT^\pm =  [\ub\, u]^{(1)} \,  [\ub\,u]^{(2)} 
\lb  \cI_{DD}^\pm + \frac{q^2}{2m^2}\, \cI_{DB}^{(w)\pm} 
+ \frac{q^4}{16 m^4}\, \cI_{BB}^{(w)\pm} \rb 
\nn\\[2mm]
&& - \frac{i}{2m} \lc [\ub\, u]^{(1)}\, [\ub \, \s_{\m\l}(p' \sm p)^\m \, u]^{(2)}
\sm   [\ub \, \s_{\m\l}(p' \sm p)^\m \, u]^{(1)}\,  [\ub\, u]^{(2)} \rc
\nn\\[2mm]
&& \times 
\frac{z^\l}{2m} \lb  \cI_{DB}^{(w)\pm} + \cI_{DB}^{(z)\pm} 
+ \frac{q^2}{4m^2}\, \cI_{BB}^{(w)\pm}\rb   
\nn\\[2mm]
&& - \frac{1}{4m^2}\,  [\ub \, \s_{\m\l}(p' \sm p)^\m \, u]^{(1)}\,  
[\ub \, \s_{\n\r}(p' \sm p)^\n \, u]^{(2)}
\lb g^{\l\r}\, \cI_{BB}^{(g)\pm} 
+ \frac{z^\l z^\r}{4m^2} (-\cI_{BB}^{(w)\pm} + \cI_{BB}^{(z)\pm}) \rb \;.
\label{3.10}
\eea

This result can be recast in a form similar to eq.(\ref{3.2}), by using the spinor 
matrix elements given in appendix B.
One finds

\bea
&& \cT^\pm = \cN^2 \, \lc  \cI_{DD}^\pm \,
\lb 2m (m \sp \cE/2)  \sm 2 (\bq_r \cd \bQ_\r)^2 / 3 \cE^2  + \bq_r^2/2 
- i \, \bsig^{(1)} \cd \bq_r \st (\bQ_r \sp \bQ_\r /\sqrt{3}) \rb \right.
\nn\\[2mm]
&& \left. \lb 2m (m \sp \cE/2)  \sm 2 (\bq_r \cd \bQ_\r)^2 / 3 \cE^2  + \bq_r^2/2 
- i \, \bsig^{(2)} \cd \bq_r \st (\bQ_r \sm \bQ_\r /\sqrt{3}) \rb + \cdots \rc \;,
\label{3.11}
\eea

\ni
and its full drift content becomes explicit.
However, for the sake of simplicity, we remain in the framework of the linear gradient 
approximation.
Using eqs.(\ref{a.6}-\ref{a.8}), one learns that, in this case, 
the variables $W^2$, $z^2$ and $q^2$ do not depart from their CM values 
and the only sources of drift corrections are the spin functions. 
The results of appendix B yield the following non-relativistic amplitude

\beq
t_2^\pm  =  \left. t_2^\pm\rb_{cm}  + \frac{\bO_D}{m^2}\, t_D^\pm \;,
\label{3.12}
\eeq

\ni
where the drift operator $\bO_D$ is given by

\beq
\bO_D = i\, (\bsig^{(1)} \sm \bsig^{(2)}) \cd \bq_r \st \bQ_\r /2\sqrt{3}
\label{3.13}
\eeq

\ni
and the profile functions read

\beq
t_D^\pm  = -  \, \frac{m}{e} 
\lc \frac{4m^2}{\l^2} \lp 1 \sp \frac{\bq_r^2}{\l^2}\rp \lb \cI_{DD}^\pm  
- \frac{\bq_r^2}{2m^2} \, \cI_{DB}^{(w)\pm}
+ \frac{\bq_r^4}{16m^4}\, \cI_{BB}^{(w)\pm} \rb 
+ \frac{e\;\bq_r^2}{m\;\l^2}\, \cI_{BB}^{(g)\pm}    \rc \;,
\label{3.14}
\eeq

\ni
with $\l^2 = 4m (e \sp m)$ and $e=\sqrt{m^2 \sp \bq_r^2 \sp 4 \bQ_r^2}$.
This result is fully model independent, since it springs directly from Lorentz covariance 
and is constrained just 
by the linear gradient approximation.
The profile functions $t_D^\pm$ do not coincide with any other components of the $TPEP$ which,
in the same approximation, are given by\cite{HR}

\bea
&& t_C^\pm = \frac{m}{e} \lc \lp 1 \sp \frac{\bq_r^2}{\l^2} \rp^2
\lb \cI_{DD}^\pm - \frac{\bq_r^2}{2m^2}\, \cI_{DB}^{(w)\pm}
+ \frac{\bq_r^4}{16m^4} \, \cI_{BB}^{(w)\pm} \rb 
+\frac{\bq_r^4}{16m^4} \, \cI_{BB}^{(g)\pm}\rc \;,
\label{3.15}\\[2mm]
&& t_{LS}^\pm = \frac{m}{e} \lc \lp 1 \sp \frac{\bq_r^2}{\l^2} \rp 
\lb- \frac{4m^2}{\l^2} \, \cI_{DD}^\pm
+ \lp1 \sp \frac{2 \bq_r^2}{\l^2}\rp \, \cI_{DB}^{(w)\pm} +  \cI_{DB}^{(z)\pm}
- \frac{\bq_r^2}{4m^2} \lp 1 \sp  \frac{\bq_r^2}{\l^2}\rp \, \cI_{BB}^{(w)\pm}\rb \right.
\nn\\[2mm]
&& \left. \;\;\;\;\;\;\;\; 
- \frac{\bq_r^2}{4m^2} \lp 1 \sp \frac{4m^2}{\l^2} \rp \, \cI_{BB}^{(g)\pm}\rc \;,
\label{3.16}\\[2mm]
&& t_{T}^\pm = t_{SS}^\pm/2 = \frac{m}{e} \lc -\frac{1}{12}\, \cI_{BB}^{(g)\pm}\rc  \;.
\label{3.17}
\eea

We consider here the expansion of the $TPEP$ to $\cO(q^4)$, using eqs.(\ref{2.9}) and (\ref{3.12}), 
which requires $t_C^\pm \!\! \rar\!\! \cO(q^4)$ and 
$\{ t_{LS}^\pm, t_T^\pm, t_{SS}^\pm, t_D^\pm  \} \!\!\rar\!\! \cO(q^{2})$. 
The expansion of the various profile functions is performed using the results
$\{\cI_{BB}^{(g)+}, \cI_{BB}^{(w)+}, \cI_{BB}^{(z)+}\} \!\! \rar \!\! \cO(q^0)$,
$\{\cI_{DB}^{(w)+}, \cI_{DB}^{(z)+}, \cI_{DB}^{(w)-}, \cI_{DB}^{(z)-}, \cI_{BB}^{(g)-}\} 
\!\! \rar \!\! \cO(q^1)$,
$\{\cI_{DD}^- \} \!\! \rar \!\! \cO(q^2)$,
$\{\cI_{DD}^+ \} \!\! \rar \!\! \cO(q^3)$, 
$\{\cI_{BB}^{(w)-}, \cI_{BB}^{(z)-} \} \sim 0$, 
and one finds

\bea
&& t_D^+  = -  \, \frac{m}{e} 
\lc  \frac{\bq_r^2}{8 m^2}\, \cI_{BB}^{(g)+}  
+ \frac{1}{2} \lb \cI_{DD}^+  - \frac{\bq_r^2}{2m^2} \, \cI_{DB}^{(w)+}\rb   \rc
\rar \lc \cO(q^2) \sp \lb \cO(q^3) \rb \rc \;,
\label{3.18}\\[2mm]
&& t_D^-  = -  \, \frac{m}{e} 
\lc \frac{1}{2}  \cI_{DD}^- \rc
\rar \lc \cO(q^2) \rc  \;.
\label{3.19}
\eea

In the expression for $t_D^+$, the term within square brackets is $\cO(q^3)$.
Nevertheless, we have kept it, for it is anomalously large.
Considering comparable terms in eq.(\ref{3.15}), one writes

\bea
&& t_C^+ = \frac{m}{e} \lc \cI_{DD}^+ - \frac{\bq_r^2}{2m^2}\, \cI_{DB}^{(w)+} \rc 
\rar \lc \cO(q^3) \rc \;,
\label{3.20}\\[2mm]
&& t_{LS}^+ = \frac{m}{e} \lc \cI_{DB}^{(w)+} +  \cI_{DB}^{(z)+}
- \lb \frac{\bq_r^2}{4m^2} \lp  \cI_{BB}^{(w)+} + \frac{3}{2}\, \cI_{BB}^{(g)+} \rp \rb \rc
\rar \lc \cO(q) \sp \lb \cO(q^2) \rb \rc \;,
\label{3.21}\\[2mm]
&& t_C^- = \frac{m}{e} \lc  \cI_{DD}^- \rc
\rar \lc \cO(q^2) \rc  \;,
\label{3.22}\\[2mm]
&& t_{LS}^- = \frac{m}{e} \lc \cI_{DB}^{(w)-} +  \cI_{DB}^{(z)-}  
- \lb \frac{1}{2} \, \cI_{DD}^- \rb \rc
\rar \lc \cO(q) + \lb \cO(q^2) \rb \rc \;.
\label{3.23}
\eea

These results show that the drift potential has little affinity with the spin orbit term and, 
at the chiral order considered, can be written as 

\bea
&& t_D^+  = \frac{3\,\bq_r^2}{4\, m^2}\, t_{SS}^+  - \frac{1}{2}\, t_C^+ \;,
\label{3.24}\\[2mm]
&& t_D^-  = - \frac{1}{2}\, t_C^- \;.
\label{3.25}
\eea

The Fourier transform of eq.(\ref{3.12}) yields the configuration space structure

\beq
V(r)^\pm  =  \left. V(r)^\pm \rb_{cm} + V_D^\pm \, \O_{D} \;, 
\label{3.26}
\eeq

\ni
with 

\bea
&& \O_D = \frac{1}{4\sqrt{3}}\, (\bsig^{(1)}\sm \bsig^{(2)}) \cd \br \st 
(-i \bnb^{^{\!\!\!\!\!\!\!\!^\leftrightarrow}}_\r) \;,
\label{3.27}\\[2mm]
&& V_D^{\pm}(r)  =  \frac{\mu^2}{m^2}\,\frac{1}{x}\, \frac{d}{d x}\,U_D^{\pm}(x)\,,
\label{3.28}\\[2mm]
&& U_D^\pm(x) =  - \int \frac{d \bq_r}{(2\pi)^3}\, e^{i\,\bq_r \cdot \br }\; t_D^\pm(q_r) \;.
\label{3.29}
\eea

The presence of the operator 
\vspace{-5mm}

\beq
\bnb^{^{\!\!\!\!\!\!\!\!^\leftrightarrow}}_\r
= \bnb^{^{\!\!\!\!\!\!\!\!^\rightarrow}}_\r
- \bnb^{^{\!\!\!\!\!\!\!\!^\leftarrow}}_\r
\label{3.30}
\eeq

\ni
in eq.(\ref{3.27}) ensures that results are symmetric under the exchange of 
initial and final states.
Using results (\ref{3.24}) and (\ref{3.25}), one has  

\bea
&& U_D^+ = \frac{3}{4}\, V_{SS}^+ - \frac{1}{2}\, V_C^+ \;,
\label{3.31}\\[2mm]
&& U_D^- = -\frac{1}{2}\, V_C^- \;.
\label{3.32}
\eea

\begin{figure}[H]
\begin{center}
\vspace{40mm}
\includegraphics[width=1.0\columnwidth,angle=0]{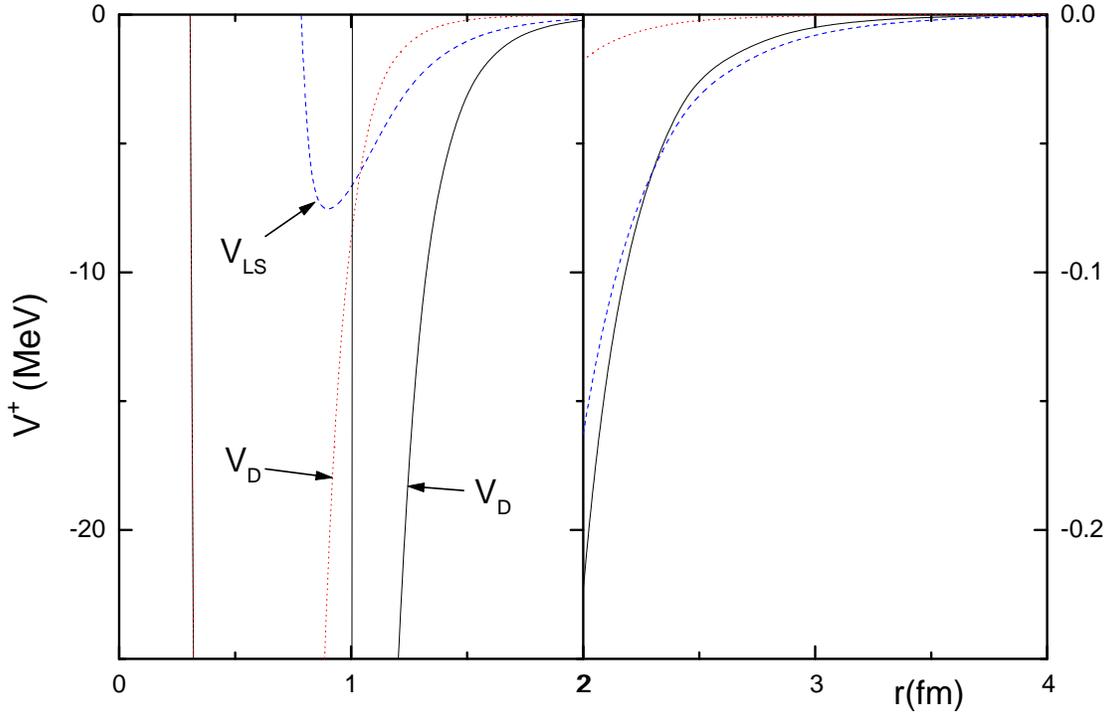}
\vspace{-75mm}
\caption{(Color online) Isospin even drift (full and dotted lines) 
and spin-orbit (dashed line) potentials;
the dotted line is $\cO(q^2)$ whereas the full one is $\cO(q^2)+\cO(q^3)$.} 
\label{F1}
\end{center}      
\end{figure}

\begin{figure}[H]
\begin{center}
\vspace{30mm}
\includegraphics[width=1.0\columnwidth,angle=0]{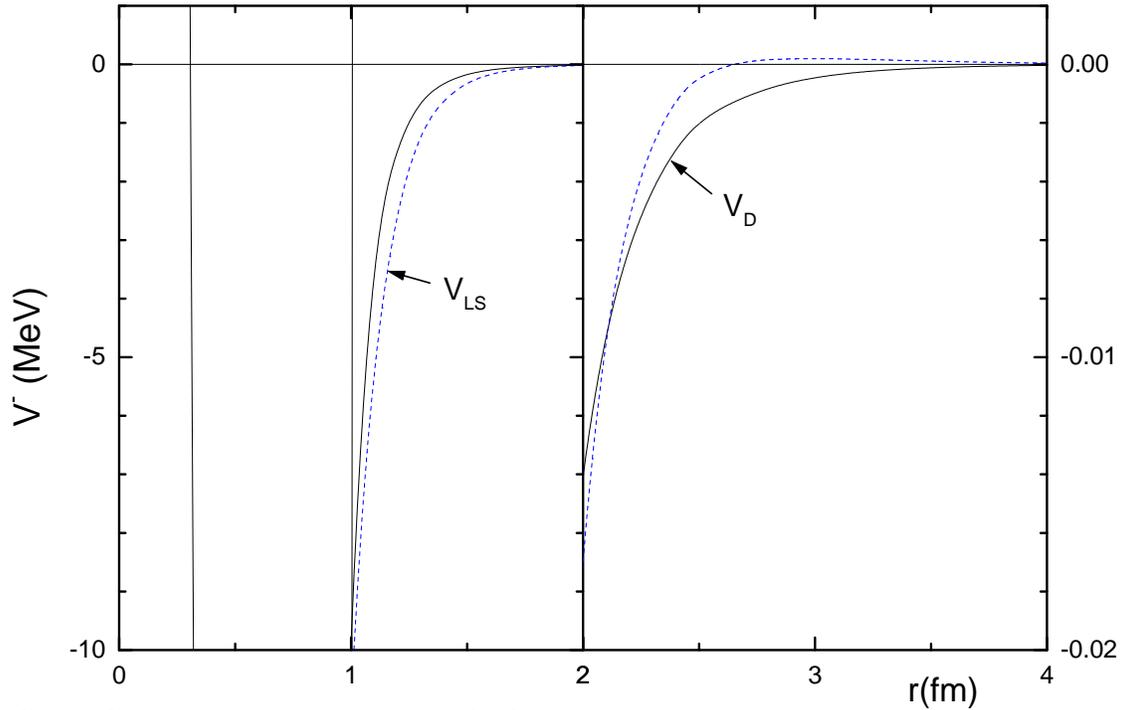}
\vspace{-75mm}
\caption{(Color online) Isospin odd drift (full line) 
and spin-orbit (dashed line) potentials.} 
\label{F2}
\end{center}      
\end{figure}

In figs.\ref{F1} and \ref{F2} we display the profile functions for the drift
and spin-orbit potentials, derived from our $\cO(q^4)$ expansion of the $TPEP$\cite{HR,HRR}.
These results do not include short range effects and cannot be trusted for $r< 1$fm.
We recall that both components of the force are multiplied by $\cO(q^2)$ spin operators
and hence we need to keep just $\cO(q^2)$ terms in $V_D$.
As shown in eqs.(\ref{3.18}-\ref{3.23}), in principle one should have
$V_{LS}^\pm \sim \cO(q) > V_D^\pm \sim \cO(q^2)$.
These expectations are confirmed in the figures, provided one uses the $\cO(q^2)$ dotted
curve for $V_D^+$.
However, when the $\cO(q^3)$ term associated with the central potential is kept,
one has a complete subversion of the expected chiral hierarchy,
associated with the prediction of a rather large effect in the isospin even channel.

In order to produce a feeling for the role of drift interactions in trinuclei, we note
that their ground states contain $S$, $P$ and $D$ waves, but they are heavily dominated by the 
principal $S$ component, which is fully symmetric under the exchange of nucleon coordinates.
Using the notation of ref.\cite{RI}, we write

\beq
| S \ra = S(\br, \bro) \; \Gamma_{1/2\,i}^{1/2 \, \m}\;,
\label{3.33}
\eeq

\ni
where 

\beq
\Gamma_{1/2\,i}^{1/2 \, \m}
= \frac{1}{\sqrt{2}}\lb |m^- \m \ra_S \, |m^+ i\ra_I  - |m^+ \m \ra_S \, |m^- i \ra_I \rb 
\label{3.34}
\eeq

\ni
is the totally antisymmetric spin-isospin $=(1/2,1/2)$ wave function with third components 
$\m$ and $i$, whereas  $|m^+ \ra$ and  $|m^- \ra$ represent respectively even and odd
mixed symmetry states under permutation of particles 1 and 2.
The leading term of the function $S(\br, \bro)$ is known\cite{Ballot} to depend just 
on the hyper-radius $\xi \equiv \sqrt{\br^2 \sp \bro^2}$ and hence the most important 
coupling introduced by the drift potential is associated with the structure

\bea
&& \O_D \, |S \ra \sim \O_D \,  S(\xi) \; \Gamma_{1/2\,i}^{1/2 \, \m}
= \frac{2 \p \, r\, \rho}{3\sqrt{3}\, \xi} \, \frac{\dr S(\xi)}{\dr \,\xi}\,
\lc \lc - \lb \lb Y_1(\hat{\br}) \otimes Y_1(\hat{\bro}) \rb_1 
\otimes |m^+ \ra_S \rb_{1/2}^\m \right. \right.
\nn\\[2mm]
&&\left. \left.
+ \sqrt{2}\, \lb \lb Y_1(\hat{\br}) \otimes Y_1(\hat{\bro}) \rb_1 
\otimes |s \ra_S \rb_{1/2}^\m \rc |m^+\, i \ra_I 
+ \lb \lb Y_1(\hat{\br}) \otimes Y_1(\hat{\bro}) \rb_1 
\otimes |m^- \ra_S \rb_{1/2}^\m \; |m^-\, i \ra_I  \rc \;,
\label{3.35} 
\eea 

\ni
$|s \ra_S$ being the spin $3/2$ state.
This result indicates that the drift potential enhances the role of $P$ waves in 
trinuclei, as one might have guessed
directly from eq.(\ref{3.27}).

\section{summary}

In nuclei containing three or more nucleons, the center of mass of a two-body subsystem
is allowed to drift.
This kind of movement does affect the forms of both one- and two-pion exchange contributions
and gives rise to important non-local corrections to the potential.
As interactions of this type are difficult to be dealt with in configuration space, 
we have restricted ourselves to the simplest possible non-local operators, 
proportional to single gradients acting on the wave function, which arise in two-pion processes.
Using a relativistic chiral expansion of the two-pion exchange $NN$ potential to $\cO(q^4)$ 
derived previously, we have shown, in a model independent way,
that the profile functions of the drift corrections do not coincide with none of its
components.
The spin dependence of the drift term is implemented by the operator

\bea
&& \O_D = \frac{1}{4\sqrt{3}}\, (\bsig^{(1)}\sm \bsig^{(2)}) \cd \br \st
(-i \bnb^{^{\!\!\!\!\!\!\!\!^\leftrightarrow}}_\r) \;,
\nn
\eea

\ni
where $\br$ and $\bro$ are Jacobi coordinates associated with two and three bodies.
This structure promotes couplings between $S$ and $P$ waves, enhancing the role 
of the latter in trinuclei.

As far as chiral symmetry is concerned, drift corrections begin at $\cO(q^4)$ and, 
in principle, should be smaller than spin-orbit terms, which begin at $\cO(q^3)$.
However, in the isospin even channel, the same dynamical contribution 
that makes the its $\cO(q^3)$ central component to be larger than the $\cO(q^2)$
odd counterpart subverts the expected chiral hierarchy and gives rise to the prediction of 
important drift effects.

\begin{acknowledgments}
I would like to thank Alejandro Kievsky for stressing the importance of having interactions
which are symmetric under exchanges between initial and final states, and
Renato Higa for supplying his numerical profile functions for the 
chiral two-pion exchange potential.
\end{acknowledgments}

\appendix
\section{kinematics}

The conventions used here are the same as in ref.\cite{HR}.
The initial and final nucleon momenta are denoted by $p$ and $p'$ and we define the variables

\bea
&& W =  p_1+p_2 = p'_1+p'_2 \;,
\label{a.1}\\
&& z  = [(p_1+p'_1) - (p_2+p'_2) ]/2 \;,
\label{a.2}\\
&& q = p'_1-p_1= p_2-p'_2 \;,
\label{a.3}
\eea

The interacting nucleons are assumed to be on shell and the following constraints hold

\bea
&& m^2 =  (W^2+z^2+q^2)/4\;,
\label{a.4}\\
&&W\cd z = W\cd q = z\cd q = 0 \;.
\label{a.5}
\eea

Using the Jacobi variables defined in eq.(\ref{2.3}), one has

\bea
&& W = (\cE, - 2 \bQ_\r/ \sqrt{3} )\;,
\label{a.6}\\[2mm]
&& z = (4\, \bQ_r \cd \bQ_\r /\cE \sqrt{3}, - 2\,\bQ_r) \;,
\label{a.7}\\[2mm]
&& q = (2\, \bq_r \cd \bQ_\r/ \cE \sqrt{3},  -\bq_r) \;.
\label{a.8}
\eea

\ni
where $\cE$ is the total energy of the two-body system.
If there were no drift, this energy would be written in terms of the single particle 
CM energy $e$ as  

\beq
\cE_{cm} = 2 \, e = 2 \, \sqrt{m^2 \sp \bq_r^2 \sp 4 \bQ_r^2} \;.
\label{a.9}
\eeq

Explicit calculation yields

\beq
\cE^2 + (4/3) \,  \lb (\bq_r \cd \bQ_\r)^2/ \cE^2 + 4\, (\bQ_r \cd \bQ_\r)^2/ \cE^2 - \bQ_\r^2 \rb 
= 4\, e^2  
\label{a.10}
\eeq

\ni
and hence, in the linear gradient approximation, 

\beq
\cE \lga 2\,e\;.
\label{a.11}
\eeq

\section{spin operators}

We present here the changes induced in  the spin operators due the drift of the two-body CM.
With our conventions, we write 

\bea
&& [\ub \, \Gamma \, u]^{(i)}
= \lc \cN \; \chi^\dagger 
\lb E' \sp m, - \bsig \cd \bp' \rb  
\lb \begin{array}{cc}
\; \cdot \; & \; \cdot \; \\
\; \cdot \; & \; \cdot \;
\end{array} \rb 
\lb \begin{array}{c}
E \sp m \\
\bsig \cd \bp
\end{array}\rb \, \chi \rc^{(i)}  \, ,
\label{b.1}\\[2mm]
&& \cN = 1 / \sqrt{(E' \sp m) (E \sp m)} \, ,
\label{b.2}
\eea

\ni
for a generic Dirac matrix $\Gamma$.
We display results for nucleon 1 and those corresponding to nucleon 2 are obtained by making 
$\bq_r \rar -\bq_r$ 
and $\bQ_r  \rar \bQ_r$.
For the normalization, one has 

\bea
&& \cN = \lb (m \sp \cE/2)^2 + 4 (m \sp \cE/2) \, \bQ_r \cd \bQ_\r / \sqrt{3}\, \cE
+ [4 (\bQ_r \cd \bQ_\r)^2 \sm (\bq_r \cd \bQ_\r)^2]/ 3\,\cE^2 \rb^{-1/2}
\nn\\[2mm]
&&  \;\;\;\;\; \lga 1/(m \sp e) \, ,
\label{b.3}
\eea

\ni
where the last equality corresponds to the linear gradient approximation.

The $OPEP$, eq.(\ref{3.1}),  is based on the function 

\bea
[\ub \, \g_5 \, u]^{(1)}\! \!&=&\! \cN \; \chi^\dagger 
\lc [m \sp \cE/2 + 2\, \bQ_r \cd \bQ_\r / \sqrt{3}\, \cE ] \, \bsig^{(1)}\cd \bq_r
- 2\, \bq_r \cd \bQ_\r \, \bsig^{(1)}\cd (\bQ_r \sp \bQ_\r / \sqrt{3}) / \sqrt{3}\, \cE \rc \chi
\nn\\[2mm]
\!& \lga &\! [\ub \, \g_5 \, u]_{cm}^{(1)} 
= \chi^\dagger\,   [\bsig^{(1)} \cd \bq_r ]\, \chi \;.
\label{b.4}
\eea

The expression for the $TPEP$ is given by eq.(\ref{3.9}) and employs the operators

\bea
&&  [\ub(\bp')\;u(\bp)]^{(1)} 
= \lc \cN \, \chi^\dagger \lb 2m (m \sp \cE/2)  - 2 (\bq_r \cd \bQ_\r)^2 / 3 \cE^2  + \bq_r^2/2 
- i \, \bsig \cd \bq_r \st (\bQ_r \sp \bQ_\r /\sqrt{3}) \rb \chi \rc^{(1)}
\nn\\[2mm]
&& \;\;\;\;\;  \lga  [\ub(\bp')\;u(\bp)]_{cm}^{(1)} 
- \lc \chi^\dagger \lb \frac{i}{(e \sp m)} \, 
\bsig \cd \bq_r \st \bQ_\r /\sqrt{3} \rb \chi \rc^{(1)}\;,
\label{b.5}\\[2mm]
&& [ \frac{i}{2m} \ub(\bp')\;\s_{\m 0}(p' \sm p)^\m\; u(\bp) ]^{(1)} 
= \lc (\cN/2m) \, \chi^\dagger \lb (m \sp \cE/2) [\bq_r^2 - 2 i \, 
\bsig \cd \bq_r \st (\bQ_r \sp \bQ_\r /\sqrt{3})] 
\right. \right.
\nn\\[2mm]
&& \;\;\;\;\; \left. \left. - 2 (\bq_r \cd \bQ_\r)  \; 
\bq_r \cd (\bQ_r \sp \bQ_\r /\sqrt{3})/ \sqrt{3}\, \cE \rb \chi \rc^{(1)} 
\nn\\[2mm]
&& \;\;\;\;\; \lga [ \frac{i}{2m} \ub(\bp')\;\s_{\m 0}(p' \sm p)^\m\; u(\bp) ]_{cm}^{(1)}
- \lc \chi^\dagger \lb \frac{i}{m} \, \bsig \cd \bq_r \st \bQ_\r /\sqrt{3} \rb \chi \rc^{(1)} \;,
\label{b.6}\\[2mm]
&& [\frac{i}{2m} \ub(\bp')\;\s_{\m j}(p' \sm p)^\m\; u(\bp) ]^{(1)}
= \lc \cN\; \chi^\dagger \lb (m \sp \cE /2) \; i\, \bsig \st \bq_r 
+ [- \bq_r^2 + 2i\, \bsig \cd \bq_r \st (\bQ_r \sp \bQ_\r /\sqrt{3}) \right. \right.
\nn\\[2mm]
&& \left. \left. \;\;\;\;\;
+ 4\, (\bq_r \cd \bQ_\r)^2 / 3 \cE^2 ] (\bQ_r \sp \bQ_\r /\sqrt{3})/2m 
- (\bq_r \cd \bQ_\r) \, [ \bq_r + 2 i \, 
\bsig \st (\bQ_r \sp \bQ_\r /\sqrt{3})] / \sqrt{3} \, \cE \rb \chi \rc^{(1)}
\nn\\[2mm]
&& \;\;\;\;\; \lga [ \frac{i}{2m} \ub(\bp')\;\s_{\m j}(p' \sm p)^\m\; u(\bp) ]_{cm}^{(1)}
- \lc \chi^\dagger \frac{1}{2(e \sp m)}\, \lb \bq_r^2 \, \bQ_\r /\sqrt{3}\,m 
+ (\bq_r \cd \bQ_\r) \, \bq_r /\sqrt{3}\,e \rb_j \chi \rc^{(1)}\;.
\label{b.7}
\eea

These results allow one to write 

\bea
&& \lc [\ub\;u]^{(1)}[\ub\;u]^{(2)} \rc 
\lga \lc \cdots \rc_{cm} - \lb \frac{4m}{(e \sp m)} 
+ \frac{\bq_r^2}{(e \sp m)^2} \rb \, \bO_D \;, 
\label{b.8}\\[2mm]
&& \lc - \frac{i}{2m}\, [\ub\;u]^{(1)}[\ub \; \s_{\m\l} \; (p'-p)^\m \;u]^{(2)}
- (1 \leftrightarrow 2) \rc \frac{z^\l}{2m}
\lga  \lc \cdots \rc_{cm} \frac{z^\l}{2m}\;,
\label{b.9}\\[2mm]
&& \lc - \frac{1}{4m^2} [\ub \; \s_{\m \l} (p'-p)^\m \;u]^{(1)}
[\ub \; \s_{\n\r}(p'-p)^\n \;u]^{(2)}\rc g^{\l\r}
\lga  \lc \cdots \rc_{cm} g^{\l\r} - \frac{e\, \bq_r^2}{m^2 (e \sp m)} \, \bO_D \;,
\label{b.10}\\[2mm]
&& \lc - \frac{1}{4m^2} [\ub \; \s_{\m \l} (p'-p)^\m \;u]^{(1)}
[\ub \; \s_{\n\r}(p'-p)^\n \;u]^{(2)}\rc \frac{z^\l z^\r}{4m^2}
\lga  \lc \cdots \rc_{cm}  \frac{z^\l z^\r}{4m^2} \;,
\label{b.11}
\eea

\ni
where the functions $\lc \cdots \rc_{cm}$ are given by eqs.(A32-A35) of ref.\cite{HR}
and the two-component spin operators $\bO$ are defined as  

\beq
\bO_D = i \, (\bsig^{(1)}\sm \bsig^{(2)}) \cd \bq_r \st \bQ_\r /2\sqrt{3} \;.
\label{b.12}
\eeq


\end{document}